\documentclass[twoside,leqno,twocolumn]{article}  
\usepackage{ltexpprt} 
\usepackage{algorithm}
\usepackage{algpseudocode}
\usepackage{graphicx}
\usepackage[tight]{subfigure}
\usepackage[show]{chato-notes}
\usepackage{epsfig}
\usepackage{url}
\usepackage{amsmath}
\usepackage{topcapt}
\usepackage{booktabs}
\usepackage{color}

\newcommand{\squishlist}{\begin{list}{$\bullet$}
  { \setlength{\itemsep}{0pt}
     \setlength{\parsep}{3pt}
     \setlength{\topsep}{3pt}
     \setlength{\partopsep}{0pt}
     \setlength{\leftmargin}{1.5em}
     \setlength{\labelwidth}{1em}
     \setlength{\labelsep}{0.5em} } }
\newcommand{\squishend}{
  \end{list}  }

\newcommand{\calP}{{\ensuremath{\mathcal{P}}}}

\newcommand{\skeleton}{{\sc skeleton}}

\newcommand{\brandes}{{\tt Brandes}}
\newcommand{\dijkstra}{{\tt Dijkstra}}
\newcommand{\dijkstrask}{{\tt Dijkstra\_SK}}
\newcommand{\brandessk}{{\tt Brandes\_SK}}
\newcommand{\buildskeleton}{{\tt Build\_SK}}

\newcommand{\centrality}{{\tt Centrality}}

\newcommand{\ouralgo}{{\tt Brandes++}}
\newcommand{\ouralgoall}{{\tt Brandes++All}}

\newcommand{\metis}{{\tt Metis}}
\newcommand{\graclus}{{\tt Gc}}

\newcommand{\modularity}{{\tt Mod}}

\newcommand{\asdata}{{\tt AS}}
\newcommand{\wikivote}{{\tt WikiVote}}
\newcommand{\dblp}{{\tt DBLP}}

\newcommand{\eumail}{{\tt EU}}

\newcommand{\Wsk}{W_{\mathrm{\sc sk}}}
\newcommand{\Vsk}{V_{\mathrm{\sc sk}}}
\newcommand{\Esk}{E_{\mathrm{\sc sk}}}
\newcommand{\Gsk}{G_{\mathrm{\sc sk}}}

\newcommand{\Ssc}{S_{\mathrm{sc}}} 
\newcommand{\Sdest}{S_{\mathrm{dest}} }

\newcommand{\spara}[1]{\smallskip\noindent{\bf{#1}}}
\newcommand{\mpara}[1]{\medskip\noindent{\bf{#1}}}

\newcommand{\footnoteremember}[2]{
\footnote{#2}
\newcounter{#1}
\setcounter{#1}{\value{footnote}}
}
\newcommand{\footnoterecall}[1]{
\footnotemark[\value{#1}]
}

\begin{document}

\title{A Divide-and-Conquer Algorithm for Betweenness Centrality
  \thanks{This research was supported in part by NSF awards PFI BIC
    \#1430145, SaTC Frontier \#1414119,  CPS \#1239021, CNS \#1012798, III
    \#1218437, CAREER \#1253393, IIS \#1320542 and gifts from Google and Microsoft.}}

\author{D\'{o}ra Erd\H{o}s\footnoteremember{thx:BU}{ Boston University, Boston
    MA \texttt{[edori, best, evimaria]@cs.bu.edu}} \and 
Vatche Ishakian\thanks{IBM T. J. Watson Research Center, Cambridge MA \texttt{vishaki@us.ibm.com}}
\and 
Azer Bestavros\footnoterecall{thx:BU}
\and
Evimaria Terzi\footnoterecall{thx:BU}}

\maketitle

\begin{abstract}
Given  a graph $G$ we define the betweenness centrality
of a node $v$ in $V$ as the fraction of shortest paths between all node pairs  in $V$ that
contain $v$. For this setting we
describe {\ouralgo}, a divide-and-conquer algorithm that can 
efficiently compute the exact values of  betweenness scores. 
{\ouralgo} uses {\brandes}-- the most widely-used algorithm
for betweenness computation -- as its subroutine. It achieves the notable faster running times by applying {\brandes} on significantly smaller
networks than the input graph, and many of its computations can be done in parallel.  
The degree of speedup achieved by {\ouralgo} 
depends on the community structure of the input network.  
Our experiments with real-life networks reveal 
{\ouralgo} achieves an average of
10-fold speedup over {\brandes}, while there are networks where this speedup is 75-fold.  We have made our code public to benefit the research community.

\end{abstract}

\section{Introduction}\label{sec:intro}
In 1977, Freeman~\cite{Freeman77} defined the betweenness centrality 
of a node $v$ as
the fraction of all pairwise shortest paths that go through $v$.
Since then, this measure of centrality has been used in a wide range of applications
including social, computer as well as biological networks.

A na\"ive algorithm can compute the betweenness centrality of a graph of $n$ nodes in 
$O(n^3)$ time. This running time was first improved in 2001 by 
Brandes~\cite{Brandes01} who
provided an algorithm that, for a graph of $n$ nodes and $m$ edges, does the same computation
in $O(nm + n^2\log n)$. The key behind this algorithm, which we call
{\brandes} is that it 
reuses information on shortest path segments that are shared by many nodes. 

Over the years, many algorithms have been proposed to 
improve the running and space complexity of {\brandes}. 
Although we discuss these algorithms in the next section, we point out here that 
most of them
either provide
approximate computations of betweenness via sampling~\cite{bader07approximating,brandes07centrality,geisberger08better,Riondato14},
or propose parallelization of the original computation~\cite{Bader06,Madduri09,Tan09,Edmonds10}.

The goal of our paper is to exploit the structure of the underlying graph and 
further improve this running time, while returning the exact values of betweenness scores.
We achieve this goal by designing the {\ouralgo} algorithm, which is a \emph{divide-and-conquer} algorithm and works as follows: first it partitions the graph into subgraphs and
runs some single-source shortest path computations on these  subgraphs.
Then it deploys a modified
version of \brandes\ on a sketch of the original graph to compute the betweenness
of \emph{all} nodes in the graph. The key behind the speed-up of \ouralgo\ over \brandes\
is that all computations are run 
over graphs that are significantly smaller
than the original graph.  

There are many real-life settings where there is a set $S$ of prominent nodes  in the network and only shortest paths connecting these nodes are important to the application. The original {\brandes}
algorithm can be used for this setting as well (see Section~\ref{sec:preliminaries} for details) and compute exactly the betweenness scores in time
$O(|S|m+|S|n\log n)$.  In this current work we first present {\ouralgo} as an algorithm that takes the target set $S$ as an input and computes the betweenness centrality of every node $v$ with respect to $S$ in Section~\ref{sec:algorithms}. We then elaborate on how to use the schema of Section~\ref{sec:algorithms} to compute betweenness centrality with respect to \emph{all} node pairs in Section~\ref{sec:allnodes}.

Our experiments (Section~\ref{sec:experiments}) with real-life networks suggest that 
there are networks for which {\ouralgo} can yield a 75-fold improvement over
{\brandes}. Our analysis reveals that this improvement depends
largely on the structural characteristics of the network and mostly on its community structure. 

Some other advantages of {\ouralgo} are the following: 
$(i)$ {\ouralgo} can employ all existing speedups for {\brandes}. $(ii)$ Many steps of our algorithm are easily paralellizable.
$(iii)$ Finally, we have made our code public to benefit the research community.

\section{Overview of Related Work}\label{sec:related}
Perhaps the most widely known algorithm for computing betweenness
centrality 
 is due to Ulrik Brandes~\cite{Brandes01}, who also studied extension of
 his 
algorithm to groups of nodes in
Brandes et al.~\cite{Brandes08}. 
The {\brandes} algorithm has motivated a lot of subsequent work 
that led to parallel versions of the
algorithm~\cite{Bader06,Madduri09,Tan09,Edmonds10} as well as classical algorithms that
approximate the betweenness centrality of
nodes~\cite{bader07approximating,brandes07centrality,geisberger08better} or
a very recent one~\cite{Riondato14}. 
The difference between approximation algorithms and \ouralgo\ is that in
case of the former a subset of the graph (either pivots, shortest paths,
etc. depending on the approach) is taken to \emph{estimate} the centrality
of 
all nodes in the graph. In contrary, \ouralgo\
computes the exact value for every node.  Further, any parallelism that can be exploited by {\brandes} can also be exploited by \ouralgo.

Despite the huge literature on the topic, there has been only little work on finding an improved centralized exact algorithm for computing betweenness centrality.
To the best of our knowledge, only recently 
Puzis
et al.~\cite{Puzis12} and  Sariy\"uce et al.~\cite{Catalyurek13} focus on that. 
In the former, the authors suggest two heuristics to speedup the computations. These heuristics
can be applied independent of each other. 
The first one, contracts \emph{structurally-equivalent} nodes (nodes that have identical neighborhoods) into one ``supernode". 
The second heuristic relies on finding the biconnected components of the graph and contracting
them into a new type of ``supernodes". These latter supernodes are then connected in the graph's biconnected tree.  The key observation is that if a shortest path has its 
endpoints in two different nodes of this tree then all shortest paths between them will traverse the same edges of the tree.  
 Sariy\"uce et al.~\cite{Catalyurek13} rely on these two heuristics and some  additional  observations
to further simplify the computations.

The similarity between our algorithm and the algorithms we described above
is in their divide-and-conquer nature.  
One can see the biconnected components of the graph 
as the input partition that is provided to {\ouralgo}.
However, since our algorithm works with \emph{any} input partition
it is more general and thus more flexible. 
Indicatively, we give some examples of how {\ouralgo} outperforms these two heuristics by comparing some of our experimental results to the results reported in 
~\cite{Puzis12} and~\cite{Catalyurek13}. In the former,  
we see that the biconnected component heuristic of Puzis~et al. achieves a $3.5$-times speedup on the \wikivote\ dataset. 
Our experiments with the same data show that {\ouralgo} provides a 
$78$-factor speedup. 
For the {\dblp} dataset Puzis~et al. achieve a speedup factor between $2-6$ -- depending on the sample. We achieve a factor of $7.8$. The best result on a social-network type graph in~\cite{Catalyurek13} is a factor of $7.9$ speedup while we achieve factors $78$ on \wikivote\ and $7.7$ on the \eumail\ data.

\section{Preliminaries}\label{sec:preliminaries}
We start this section by defining betweenness centrality. Then we review
some necessary previous results.

\mpara{Notation.}
Let $G(V,E,W)$ be an undirected weighted graph with 
nodes $V$, edges $E$  and  non-negative 
edge weights $W$. We denote $|V| = n$ and $|E| = m$. 

Let $u,v \in V$. The {\em distance} between $u$ and $v$ is the \emph{length} of the (weighted) 
shortest path in $G$ connecting them, we denote this by $d(u,v)$. 
We denote by $\sigma(u,v)$ the \emph{number} of shortest paths between $u$ and $v$. 
For $s,t \in V$ the value  $\sigma(s,t|v)$ denotes the number of shortest paths connecting $s$ and $t$ 
that contain $v$. 
Observe, that $\sigma$ is a symmetric function, thus $\sigma(s,t) = \sigma(t,s)$. 

The {\em dependency} of $s$ and $t$ on $v$ is the fraction of shortest paths connecting $s$ and $t$ that go through $v$, thus \[
\delta(s,t|v) = \frac{\sigma(s,t,|v)}{\sigma(s,t)}.
\]
Given the above, the {\em betweenness centrality} $C(v)$ of node $v$ can be defined as the sum of its dependencies.
\begin{equation}\label{eq:centralityall}
C(v) = \sum_{s \neq t \in V}\delta(s,t|v).
\end{equation}
Throughout the paper we use the terms betweenness, centrality and betweenness centrality interchangeably.

\spara{A na\"ive algorithm for betweenness centrality.} In order to compute
the dependencies in Eq.~\eqref{eq:centralityall} 
we need to compute $\sigma(s,t)$ and $\sigma(s,t|v)$ for every triple $s,t$
and $v$. 
Observe that $v$ is contained in a shortest path between $s$ and $t$ if and only if $d(s,t) = d(s,v) + d(v,t)$. 
If this equality holds, then any shortest path from $s$ to $t$ can be
written as the concatenation 
of a shortest path connecting $s$ and $v$ and a shortest path from $v$ to $t$. 
Hence, $\sigma(s,t|v) = \sigma(s,v)\cdot\sigma(v,t)$.  
If $P_v = \{u \in V| (u,v) \in E, d(s,v) = d(s,u) + w(u,v)\}$ is the set of parent nodes of $v$, then 
it is easy to see that 
\begin{equation}\label{eq:parentsall}\sigma(s,v) = \sum_{u \in P_v}\sigma(s,u).
\end{equation} 
We can compute $\sigma(s,v)$ for a given target $s$ and all possible 
nodes $v$ by running a weighted single source shortest paths algorithm 
(such as the \dijkstra\ algorithm) with source $s$. 
While the search tree in \dijkstra\ is built $\sigma(s,v)$ is computed by formula~\eqref{eq:centralityall}.
 The running time of \dijkstra\ is $O(m + n \log n)$ per source using a Fibonacci-heap implementation (the fastest known implementation of \dijkstra).
Finally, a na\"ive computation of the dependencies can be done as 
$$\delta(s,t|v) = \frac{\sigma(s,v)\cdot\sigma(t,v)}{\sigma(s,t)}.$$ 
Even given if all $\sigma(s,t)$ values are given, this computations
requires time equal to the number of dependencies, i.e., $O(n^3)$. 

\spara{The {\brandes} algorithm.}
Let $\delta(s|v)$ define the dependency of a node $v$ on a single target $s$ as the sum of the dependencies  containing $s$, thus \begin{equation}\label{eq:depall}\delta(s|v) = \sum_{t \in V}\delta(s,t|v). \end{equation} 
The key observation of {\brandes} is that 
for a fixed target $s$ we can compute $\delta(s|v)$ by 
traversing the shortest-paths tree found by {\dijkstra}
in the reversed order of distance to $s$
using the formula:
\begin{equation}\label{eq:brandesall}
\delta(s|u) = \sum_{v: u \in P_v}\frac{\sigma(s,u)}{\sigma(s,v)}(1+\delta(s|v)).
\end{equation}
Using this trick, the dependencies can be computed in time $O(nm)$, yielding a total
running time of $O(nm + n^2\log n)$ for {\brandes}.

\spara{Betweenness centrality for a given target set.} In many applications there is a subset of nodes  $S \subseteq V$ that is of interest to the user. We call $s \in S$ a {\em target}
node and assume 
$2\leq |S|\leq n$.
Observe now, that the na\"ive algorithm for betweenness centrality can easily be modified to compute the centralities only with respect to $S$. For this we only need to modify equation~\eqref{eq:centralityall} to sum  over nodes in the target set only, \begin{equation}\label{eq:centrality} C^S(v) = \sum_{s \neq t \in S}\delta(s,t|v).\end{equation}  Observe that for $S=V$ the definitions in equations~\eqref{eq:centralityall} and~\eqref{eq:centrality} are identical. As it will always be clear from the context whether the centrality of $v$ is computed with regard to  a target set $S$ or the entire $V$, we will omit $S$ from the notation and use $C(v)$ instead of $C^S(v)$ in this paper. Naturally we only compute $\sigma(s,v)$ for pairs where $s \in S$ and $v \in V$. 
This modified algorithm
requires time equal to the number of dependencies, that is $O(|S|^2\cdot n)$. 

To adjust the {\brandes} algorithm to the target set we again  need to modify the computations to only consider nodes in $S$. Thus the dependency $\delta(s|v)$  only takes targets $t \in S$ into consideration;
\begin{equation}\label{eq:dep}\delta(s|v) = \sum_{t \in S}\delta(s,t|v). \end{equation}
The recusive formula in equation~\eqref{eq:brandesall} also only takes target nodes into account;
\begin{equation}\label{eq:brandes}
\delta(s|u) = \sum_{v: u \in P_v}\frac{\sigma(s,u)}{\sigma(s,v)}(I_{v \in S}+\delta(s|v)).
\end{equation}
Where $I_{v \in S}$ is an indicator that is $1$ if $v \in S$ and zero otherwise. This is used to make sure that we only sum dependencies between pairs of target nodes.
Using this trick, the dependencies can be computed in time $O(|S|m)$, yielding a total
running time of $O(|S|m + |S|n\log n)$ for {\brandes}.

\section{The {\large \skeleton} Graph}\label{sec:skeleton}
In this section, we introduce the \skeleton\ of a graph $G$. The purpose of the \skeleton\ is to get a simplified representation of $G$ that still contains all information on shortest paths between nodes. 

Let
$G(V,E,W)$ be a weighted undirected graph with nodes $V$, edges $E$ and edge weights  $W: E \to [0,\infty)$. 
We also assume that we are given a partition $\calP$ of the nodes $V$  into $k$ parts: $\calP=\{P_1,\ldots ,P_k\}$
such that $\cup_{i=1}^k P_i=V$ and $P_i\cap P_j=\emptyset$ for every $i\neq
j$.  

The \skeleton\ of $G$ is defined to be a 
graph $\Gsk^\calP(\Vsk,\Esk, \Wsk)$; its nodes $\Vsk$ are a subset of
$V$. For every edge $e \in \Esk$ the function $\Wsk$ represents a pair of
weights called  the \emph{characteristic tuple} associated with $e$. All of
$\Vsk$, $\Esk$ and $\Wsk$  depend on the partition $\calP$. Whenever it is clear from the context which partition is used we drop $\calP$ from the notation and use $\Gsk$ instead of $\Gsk^\calP$. We now
proceed to explain in detail how $\Vsk$, $\Esk$ and $\Wsk$ are defined.

\spara{Supernodes:} Given $\calP$, we define $G_i$ to be the subgraph of $G$ that is spanned by the nodes in $P_i \subseteq V$, that is 
$G_i=G[P_i]$. We denote the nodes and edges of $G_i$ by $V_i$ and $E_i$ respectively. We refer to the subgraphs $G_i$ as \emph{supernodes}. Since $\calP$ is a partition, all nodes in $V$ belong to one of the supernodes $G_i$.

\begin{figure*}[t]
\hfill
\subfigure[][original graph]{
\includegraphics[scale=0.25]{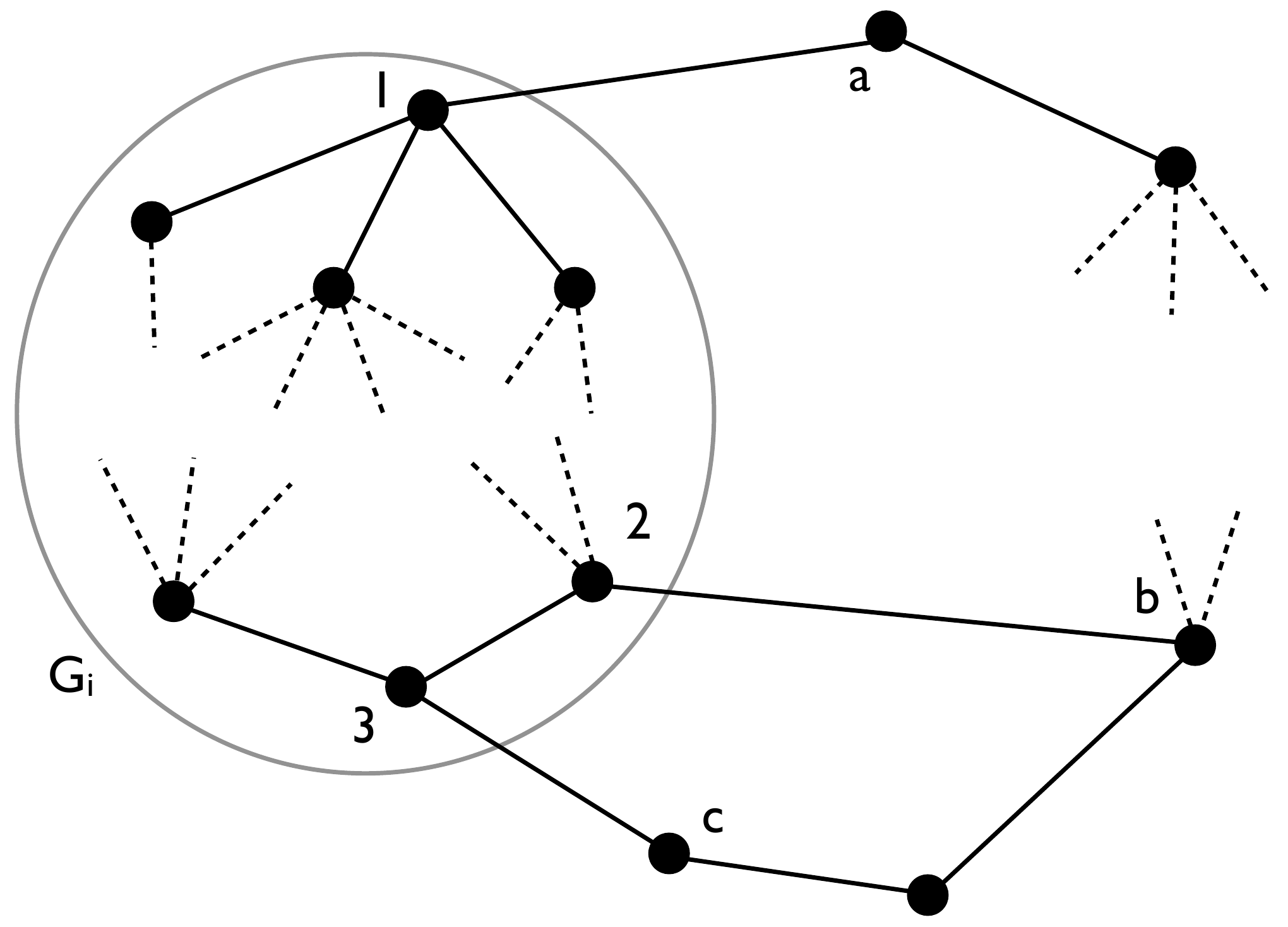}
\label{fig:G}
}\hfill
\subfigure[][\skeleton]{
\includegraphics[scale=0.25]{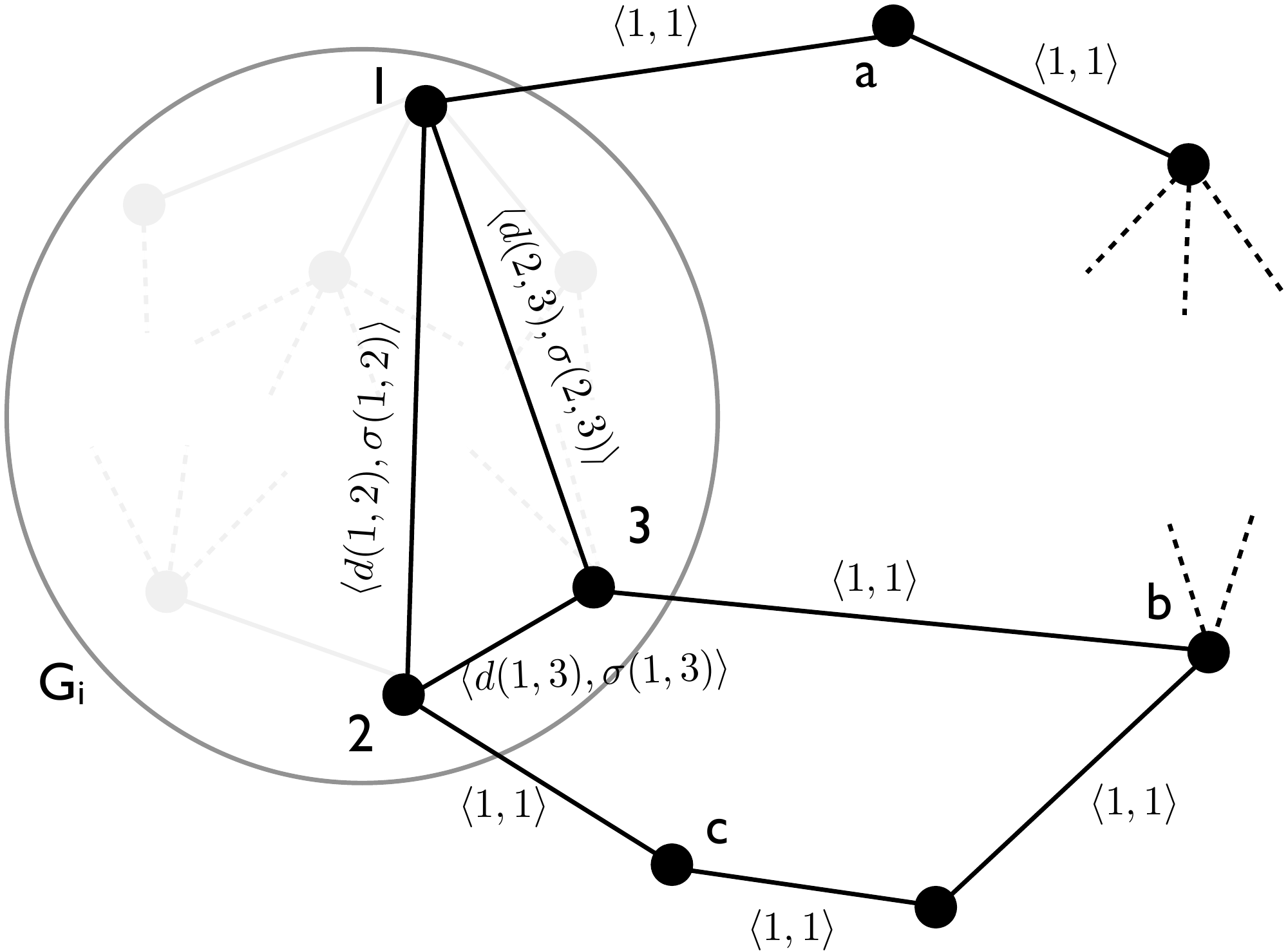}
\label{fig:sk}
}\hfill
\vspace{-0.1in}
\caption[]{\label{fig:allsk}Graph $G(V,E)$ (Figure~\ref{fig:G}) is given as input
    to \ouralgo. The nodes and edges inside the circle correspond to
    supernode $G_i$. The set of frontier nodes in $G_i$ is $F_i = \{1,2,3\}$. Supernode $G_i$ is  replaced by a clique on nodes $\{1,2,3\}$ with
characteristic tuple $\langle d_{jk},\sigma_{jk}\rangle$ on edge $(j,k)$ in
the \skeleton\ (Figure~\ref{fig:sk}).}
\vspace{-0.1in}
\end{figure*}

\spara{Nodes in the {\skeleton} ($\Vsk$):} Within every supernode
$G_i(V_i,E_i)$ there are 
some nodes $F_i\subseteq V_i$ of special significance. 
These are the nodes that have at least one edge connecting them to a node of another supernode $G_j$.
We call  $F_i$ the \emph{frontier} of  $G_i$.  In Figure~\ref{fig:G} the supernode $G_i$ consists of nodes and edges inside the large circle. The frontier of $G_i$  is $F_i = \{ 1,2,3\}$. 
 Observe that nodes $a$, $b$
and $c$ are also frontier nodes in their respective supernodes. The \emph{nodes} $\Vsk$ of the \skeleton\ consist of the union of all frontier nodes i.e., $\Vsk = \cup_{i=1}^k F_i$.

\spara{Edges in the {\skeleton} ($\Esk$):} The edges in $\Gsk$ are defined with help of the frontiers in $G$.
First, in order to see the significance of the frontier nodes, pick any two 
 target nodes $s,t \in V$.  
 Observe, that some of the shortest paths between $s$
and $t$ may pass through $G_i$. Any such path has to enter the supernode through one of the frontier nodes $f \in
F_i$ and exit through another frontier $q \in F_i$. It is easy to check, whether there are any shortest paths through $f$ and $q$; given
$d(f,q)$, there is a shortest path between $s$ and $t$
passing through $f$ and $q$  if and only if 
\begin{equation}\label{eq:d(s,t)}d(s,t) = d(s,f) +
d(f,q) + d(q,t).\end{equation}
Also the number of
paths passing through $f$ and $q$ is: \begin{equation}\label{eq:s(f,q)}\sigma_G(s,t|f,q)
= \sigma(s,f)\cdot \sigma(f,q) \cdot \sigma(q,t).\end{equation}
Recall that the nodes $\Vsk$ of the \skeleton\ are the union of all
frontiers in the supernodes. The \emph{edges} $\Esk$ serve the purpose of
representing the possible shortest paths between pairs of frontier nodes,
and as a result, the paths between pairs of target nodes in $G$.  The key observation to the definition of the \skeleton\ is, that we solely depend on the frontiers and do not need to list all possible (shortest) paths in $G$. We want to
emphasize here that in order not to double count, we only consider the
paths connecting $f$ and $q$ that do not contain any other frontier inside
the path. Paths containing more than two frontiers will be considered as
concatenations of shorter paths during computations on the
entire \skeleton. The exact details will be clear once we define the edges and some weights assigned to the edges in the following paragraphs.

$\Esk$  consists of two types of edges; first, the edges
that connect frontiers in different supernodes 
(such as edges $(1,a)$, $(2,b)$ and $(3,c)$ in Figure~\ref{fig:G}). 
We denote these edges by $R$. 
Observe that these edges are also in the original graph $G$, namely $R = E\setminus \{\cup_{i=1}^k E_i\}$.  
The second type are edges between all pairs of frontier nodes $f,q \in F_i$ within each supernode.  
To be exact, we add the edges $X_i$ of the  clique  $C_i = (F_i, X_i)$ to the {\skeleton}. 
Hence, the edges of the \skeleton\ can be defined as the union of $R$ and the cliques 
defined by the supernodes, i.e., $\Esk = R \cup \{ \cup_{i=1}^k X_i\}$.

\spara{Characteristic tuples in the \skeleton\ ($\Wsk$):} We assign 
a \emph{characteristic tuple} $\Wsk(e) = \langle \delta(e), \sigma(e) \rangle$, consisting of a \emph{weight} and a \emph{multiplicity}, to every edge $e\in\Esk$.
For edge $e(u,v)$ the weight represents the length of the shortest path between $u$ and $v$ in the
original graph; the multiplicity encodes the number of different shortest paths between these two nodes.
That is, if $e\in R$, then  $\Wsk(e) = \langle w(e),1\rangle$, where $w(e)$ is the weight of $e$ in $G$.
If $e = (f,q)$ is in $X_i$ for some $i$, then $f,q \in F_i$ are frontiers in $G_i$.
In this case $\Wsk(e) = \left\langle d(f,q),\sigma(f,q)\right\rangle$.  The values  $d(f,q)$ and
$\sigma(u,v)$  are used in Equations~\eqref{eq:d(s,t)} and~\eqref{eq:s(f,q)}. While these equations allow to compute the distance $d(s,t)$ and multiplicity $\sigma(s,t | f,q)$ between target nodes $s$ and $t$,  both values  are independent of the target nodes themselves. In fact, $d(f,q)$ and $\sigma(f,q)$ only depend and are characteristic of their supernode $G_i$.

We   compute $d(f,q)$ and $\sigma(f,q)$ by applying \dijkstra\  -- as
described in section~\ref{sec:preliminaries} --  in $G_i$ using the set of
frontiers $F_i$ as sources. We want to emphasize here, that the
characteristic tuple only represent the shortest paths between $f$ and $q$
that are entirely within the supernode $G_i$ and do not contain any other
frontier node in $F_i$. This precaution is needed to avoid double counting
paths between $f$ and $q$ that leave $G_i$ and then come back later.   To ensure this, we apply a very simple modification to the \dijkstra\ algorithm; in equation~\eqref{eq:parents} we only sum over the set of parents $P_v^{-}$ of a node that are \emph{not} frontiers themselves, thus 
\begin{align}\label{eq:frontierparents}\lefteqn{P_v^{-}}\ \ \ = & \{u \in V_i\setminus F_i\  |\\ &\  (u,v) \in E_i, d(s,v) = d(s,u) + w(u,v)\}.\nonumber\end{align} 
We refer to this modified version of \dijkstra's algorithm that is run on the supernodes as \dijkstrask. 
For recursion~\eqref{eq:s(f,q)} we also set $\sigma(f,f) = 1$.

\spara{The \skeleton:} Combining all the above, the \skeleton\ of a graph $G$ is defined by the supernodes generated by the partition $\calP$  and can be described formally as
\[\Gsk^\calP = (\Vsk,\Esk,\Wsk) = (\cup_{i=1}^k F_i, R \cup \{\cup_{i=1}^k X_i\},\Wsk)\]
Figure~\ref{fig:sk}
shows the \skeleton\ of the graph from Figure~\ref{fig:G}. The nodes in $\Gsk$ are the frontiers of $G$ and the edges are the dark edges in this picture. Edges in $R$ are for example $(1,a)$, $(2,b)$ and $(3,c)$ while edges in $X_i$ are  $(1,2)$, $(1,3)$ and $(2,3)$. 

\spara{Properties of the {\skeleton}:}
We conclude this section by comparing the number of nodes and edges of the input graph $G=(V,E,W)$ and its skeleton
$\Gsk=(\Vsk,\Esk,\Wsk)$. This comparison will facilitate the
computation of the running time of the different algorithms in the next section.

Note that $\Gsk$ has less nodes than $G$: the latter has $|V|$ nodes, while
the former has only $\left|\Vsk\right| = \sum_{i=1}^k\left|F_i\right|$. Since not all nodes in $\Gsk$ are frontier nodes,
then $|V|\leq \left|\Vsk\right|$. For the edges, the original graph has $\left|E\right|$
edges, while its skeleton has $\left|\Esk\right|=\left|E\right|-\sum_{i=1}^k\left|E_i\right|+\sum_{i=1}^k{\left|F_i\right|\choose k}$.
The relative size of $\left|E\right|$ and $\left|\Esk\right|$ depends on the partition $\calP$ and the number of
frontier nodes and edges between them it generates. 

\section{The {\large \ouralgo} Algorithm for Target Set $S$}\label{sec:algorithms}
We are now ready to describe {\ouralgo}, which  leverages the speedup that can be gained by using the \skeleton\ of a graph. At a high level \ouralgo\ consists of three main steps, first the \skeleton\ is created, then  a multipiclity-weighted version of \brandes's algorithm is run on the \skeleton. 
In the final step the centrality of all other nodes in $G$ is computed.

In this section we present $\ouralgo$ as it is applied to computing the betweenness centrality of nodes with regard to a target set $S$. It is trivial to see that the results of this section could be used to compute betweenness over all node pairs by taking $S=V$. However, as we will see, the running time for {\ouralgo}, when taking $S$ into consideration is dependent on $S$ and suboptimal compared to $\brandes$ if $|S|$ is too large.  In the next Section~\ref{sec:allnodes} we explain how to compute centrality over all node pairs, again by leveraging the {\skeleton}. We denote the version of {\ouralgo} that considers all node pairs by {\ouralgoall}.  

\mpara{The {\ouralgo} algorithm:} The pseudocode of \ouralgo\ is given in Alg.~\ref{algo:ouralgo}.
The input to this algorithm is the weighted undirected graph $G=(V,E,W)$, the set of
targets $S$ and partition $\calP$. The algorithm 
outputs the exact values of  betweenness centrality for every node in $V$.
Next we explain the details of each step.

\begin{algorithm}[ht!]
\begin{algorithmic}[1]
\Statex {\bf Input:} graph $G(V,E,W)$, targets $S$, partition $\calP=\{P_1,\ldots ,P_k\}$.
\State  $\Gsk(\Vsk,\Esk,\langle.,.\rangle) = \buildskeleton(G,\calP)$\label{ln:buildsk}
\State $\{C(G_1),\ldots,C(G_k) \} = \brandessk(\Gsk)$\label{ln:brandes}
\State $\{ C(v) |v \in V\} = \centrality(\{C(G_1),\ldots,C(G_k) \})$\label{ln:centrality}
\Statex {\bf return:} $C(v)$ for every $v \in V$
\end{algorithmic}
\caption{\label{algo:ouralgo} \ouralgo\ to  compute the exact betweenness centrality of all nodes for a target set $S$.}
\end{algorithm}

\spara{Step 1: The \buildskeleton\ algorithm:} 
  \buildskeleton\ (Alg.~\ref{algo:buildskeleton}) takes
  as input $G$ and the partition $\calP$ and outputs the \skeleton\
  $\Gsk(\Vsk,\Esk, \Wsk)$. First  it decides the set of frontiers $F_i$ in
  the supernodes (line~\ref{ln:frontiers}). Then the characteristic tuples
  $\Wsk$ are computed in every supernode by way of \dijkstrask\
(line~\ref{ln:dijkstra}). Characteristic tuples on edges $e \in R$ are
  $\langle 1,1\rangle$ by definition.

\begin{algorithm}[ht!]
\begin{algorithmic}[1]
\Statex {\bf Input:} graph $G(V,E,W)$, targets $S$, partition $\calP$.
\State Find frontiers $\{F_1,F_2,\ldots,F_k\}$\label{ln:frontiers}
\For{$i = 1$ to $k$}
\State $\{ \langle d(f,q),\sigma(f,q) \rangle \ |\ \mathrm{for\ all\ }f,q, \in F_i\} = \dijkstrask(F_i)$\label{ln:dijkstra}
\EndFor
\Statex {\bf return:} \skeleton\ $\Gsk(\Vsk,\Esk,\Wsk)$
\end{algorithmic}
\caption{\label{algo:buildskeleton} \buildskeleton\ algorithm to create the \skeleton\ of $G$.}
\end{algorithm}

\emph{Running time:} The frontier sets $F_i$ of each supernode can be found in $O(|E|)$ time
as it requires to check for every node whether they have a neighbor in
another supernode. \dijkstrask\ has running time identical to the
traditional \dijkstra\ algorithm, that is $O(|F_i|(|E_i| + |V_i|\log |V_i|))$.

\emph{Target nodes in the {\skeleton}:} Note
that since we need to know the shortest paths for every target node $s\in S$, 
we treat the nodes in $S$ specially. More specifically,
given the input partition $\calP$, we remove 
all targets from their respective parts and add them as 
singletons. Thus, we use the partition $\calP' = \{P_1\setminus S,
P_2\setminus S,\ldots,P_k\setminus S,\cup_{s \in S}\{s\}\}$. 
Assuming that the number of target nodes is relatively small compared to the total number
of nodes in the network, this does not 
have a significant effect on 
the running time of our algorithm.



Observe that the characteristic tuples of different supernodes are independent of each other allowing for a parallel execution of {\dijkstrask}.


\spara{Step 2: The \brandessk\ algorithm:} 
The output of \brandessk\ are the exact
betweenness centrality values for all nodes in $\Gsk$, that is all
frontiers in $G$.

Remember from Section~\ref{sec:preliminaries} that for every target node
$s \in S$ \brandes\ consists of two main steps; (1) running a single-source
shortest paths algorithm from $s$ to compute the distances $d(s,v)$ and  number of
shortest paths $\sigma(s,v)$. 
(2) traversing the BFS tree of \dijkstra\ in reverse order of discovery to
compute the dependencies $\delta(s|v)$ based on Equation~\eqref{eq:brandes}.
The only difference between \brandessk\  and \brandes\ is that we take the distances and
multiplicities on the edges of the \skeleton\ into consideration.  
This means that Equation~\eqref{eq:multiparents} is used  instead
of~\eqref{eq:parentsall}. \begin{equation}\label{eq:multiparents}\sigma(s,v) = \sum_{u \in P_v^{\sc sk}}\sigma(s,u)\sigma(u,v).\end{equation}
Here $P_v^{\sc sk} = \{u \in \Vsk| (u,v) \in \Esk, d(s,v) = d(s,u) + d(u,v)\}$ is the set of parent nodes of $v$ in $\Gsk$.
Observe that $\sigma(s,v)$  in Equation~\eqref{eq:multiparents} is  the
multiplicity of shortest paths between $s$ and $v$ both in $\Gsk$ as well as in $G$. That is why we do not use subscripts (such as $\sigma_{\sc sk} (s,v)$) in the above formula.


In the second step the dependencies of nodes in $\Gsk$ are computed by 
applying Equation~\eqref{eq:multibrandes} -- which is the counterpart of
Eq.~\eqref{eq:brandes} that takes multiplicities into account -- to  the reverse order traversal of the BFS
tree. 
\begin{equation}\label{eq:multibrandes}\delta(s|u) = \sum_{v: u \in P_v}m(u,v)\cdot\frac{\sigma(s,u)}{\sigma(s,v)}(I_{v \in S}+\delta(s|v))\end{equation}



\emph{Running time:} \brandes\ and \brandessk\ have the same computational
complexity but are applied to different graphs ($G$ and $\Gsk$
respectively). Hence we get that \brandessk\ on the skeleton runs in $O(|S|\Esk + |S|\Vsk \log \Vsk)$ time.
 If we express the same running time in terms of the frontier nodes we get 
{\small
 \[O\left(|S|(|R| + \sum_{i=1}^k{|F_i| \choose 2}) + |S|\left(\sum_{i=1}^k|F_i|\right)\log \left(\sum_{i=1}^k|F_i|\right)\right).\]
}



\spara{Step 3: The \centrality\ algorithm:}
In the last step of {\ouralgo}, the centrality values of all remaining
nodes  $v \in V_i\setminus F_i$ in $G$ are computed.
Let us focus on supernode
$G_i$; for any node $v \in V_i\setminus F_i$ and $s\in S$ there exists a frontier 
$f \in F_i$ such that there
exists a shortest path from $s$ to $f$ containing $v$. 
Using Equation~\eqref{eq:multibrandes}, we can compute
the
dependency $\delta(s|v)$ as follows: 
\begin{equation}\label{eq:localcentrality}
\delta(s|v) = \sum_{f \in F_i} \frac{\sigma(s,v)}{\sigma(s,f)}\sigma(v,f)\left(I_{v \in
S}+\delta(s|f)\right).
\end{equation}
Then, the centrality of $v$ is  $C(v) = \sum_{s
\in S} \delta(s|v)$.

To determine
whether $v$ is contained in a path from $s$ to $f$ we need to remember the information
$d(f,v)$ for $v$ and every frontier $f \in F_i$. This value is actually
computed during the \buildskeleton\ phase of \ouralgo. Hence with
additional use of space but without increasing the running time of the
algorithm we can make use of it. At the same time with $d(f,v)$ the
multiplicity $\sigma(f,v)$ is also computed. 

\emph{Space complexity:} The \centrality\ algorithm takes two values -- $d(f,v)$ and $\sigma(f,v)$ -- for every pair $v \in V_i\setminus F_i$ and $f \in F_i$. This results in storing a total of $\sum_{i=1}^k\left(|F_i||V_i\setminus F_i|\right)$ values for the \skeleton.

\emph{Running time:} Since we do not need to allocate any additional
time for computing $d(f,v)$ and $\sigma(f,v)$  computing
Equation~\eqref{eq:localcentrality} takes $O(|F_i|)$ time for every $v \in
V_i\setminus F_i$. Hence, summing over all supernodes we get that the
running time of \centrality\ is 
 $O\left(\sum_{i=1}^k|F_i||V_i\setminus F_i|\right)$.
 
 \spara{Running time of \ouralgo:}  The total time that \ouralgo\  takes is the combination of  time required for steps 1,2 and 3. The asymptotic running time is a function of the number of nodes and edges in each supernode, the number of frontier nodes per supernode and  the size of the {\skeleton}. To give some intuition, assume that all supernodes have approximately $\frac{n}{k}$ nodes with at most half of the nodes being frontiers in each supernode. Further, assume that $R \leq \frac{m}{2}$.  Substituting these values into steps 1--3, we get that for a partition of size $k$ \ouralgo\ is order of $k$-times faster than {\brandes}. While these assumptions are not necessarily true, they  give some insight on how \ouralgo\ works.   For $k=1$ (thus when there is no partition) the running times of \ouralgo\ and \brandes\ are identical while for larger values of $k$ the computational speedup is much more significant.

\section{{\large \ouralgoall} for All Pairs of Nodes}\label{sec:allnodes}
The concept of the {\skeleton} graph is also suitable to compute betweenness centrality over all pairs of nodes in $G$. In this section we present this version of {\ouralgo}, which we denote by {\ouralgoall}.

\mpara{The {\ouralgoall} algorithm:} The high level structure of {\ouralgoall}, shown in Algorithm~\ref{algo:ouralgoall}, is very similar to Algorithm~\ref{algo:ouralgo} presented in Section~\ref{sec:algorithms}.  
\begin{algorithm}[ht!]
\begin{algorithmic}[1]
\Statex {\bf Input:} graph $G(V,E,W)$, partition $\calP=\{P_1,\ldots ,P_k\}$.
\State $C(v) = 0, \forall v \in V$
\State  $\Gsk(\Vsk,\Esk,\langle.,.\rangle) = \buildskeleton(G,\calP)$\label{ln:buildskall}
\For{$i,j = 1,2,\ldots k$}
\State $\Ssc= P_i$, $\Sdest = P_j$
\State $\{C_{ij}(G_1),\ldots,C_{ij}(G_k) \} =$\\$\brandessk(\Gsk,\Ssc,\Sdest)$\label{ln:brandesall}
\State $\{ C_{ij}(v)\}$ $ =\centrality(\{C_{ij}(G_1),\ldots,C_{ij}(G_k) \})$\label{ln:centralityall}
\For{$v \in V$}
\State $C(v) += C_{ij}(v)$
\EndFor
\EndFor
\Statex {\bf return:} $C(v)$ for every $v \in V$
\end{algorithmic}
\caption{\label{algo:ouralgoall} \ouralgoall\ to  compute the exact betweenness centrality of all nodes in $V$.}
\end{algorithm} 
{\ouralgoall} takes as input the graph $G(V,E,W)$ and the partition $\calP=\{P_1,\ldots ,P_k\}$.    We create the supernodes $G_1,G_2,\ldots,G_k$ the same way as before with help of the {\buildskeleton} algorithm (Line~\ref{ln:buildskall}). We set the centrality $C(v)$ for every node $v$ initially to $0$.  The idea is to iterate over all pairs of supernodes $G_i$ and $G_j$ and compute the centrality $C_{ij}(v)$ of nodes when we only consider shortest paths that have a node in $G_i$ as their source and a node in $G_j$ as their destination. We compute the centrality of $v$ as the sum $$C(v) = \sum_{i,j = 1\ldots k}C_{ij}(v).$$ Since $\calP$ is a partition of $G$ and we iterate over all $i,j$ pairs (including the case $i=j$), this way we consider all shortest paths in $G$. We now discuss the steps in Algorithm~\ref{algo:ouralgoall} in detail.

\mpara{Step 1, the {\buildskeleton} algorithm:} In this first step we compute the {\skeleton} $\Gsk$ the same way as before. Note that there is no designated target set, hence $\Gsk$ will consist exactly of the nodes and edges defined by the partition $\calP$.

\mpara{Step 2, the {\brandessk} algorithm:} The version of {\brandessk} that we use here has one additional step to the algorithm described in Section~\ref{sec:algorithms}. It takes as input not only $\Gsk$ but also the set of source nodes $\Ssc$ and destination nodes $\Sdest$. First it will change $\Gsk$ by adding every node in $\Ssc$ and $\Sdest$  as singleton supernodes to the graph, the resulting {\skeleton} is denoted by $\Gsk^{ij}$. Then the old version of  $\brandessk(\Gsk^{ij})$ is run on this new {\skeleton} that is dependent on $i$ and $j$. Note that when $i=j$, then $\brandessk(\Gsk^{ii})$ is exactly the algorithm described in the previous section with $S = V_i$. If $i \neq j$, then the indicator function in Equation~\eqref{eq:brandes} is $1$ if $v \in \Sdest$ and zero otherwise.

\mpara{Step 3, the {\centrality} algorithm:} This algorithm is identical to {\centrality} described in the previous section.

\mpara{Running time of {\ouralgoall}:} The main difference in the running time between {\ouralgo} and {\ouralgoall} is that in the latter {\brandessk} is called $k^2$ times as opposed to once. But, since it is run on the same size {\skeleton} this only yields  a $k^2$-factor increase in this part of the algorithm.   Observe that the running times of {\buildskeleton} and {\centrality} did not change in this version of {\ouralgo}. For {\buildskeleton} this is trivial to see. Let $v$ be any node in $V$ and let $G_v$ be the supernode that contains $v$. To see the claim for {\centrality}, recall that  to compute $\delta(s,t|v)$ for some source $s$, destination $t$ and node $v$, we need to do a computation for all $f,g$ frontier tuples in  $F_v$ (if $v \notin F_v$). Thus, the number of required computations is the same as in Section~\ref{sec:algorithms}, not forgetting that this number is a function of the size of the target set which in this instance is $n$.


\section{Experiments}\label{sec:experiments}
In this section, we validate the performance of {\ouralgo} for a given target set $S$, with
experiments using data from a diverse set of applications.

\spara{Experimental setup:} For all  our experiments we follow the same
methodology; given the partition $\calP$, we run Steps 1--3 of \ouralgo\
(Alg.~\ref{algo:ouralgo}) using  $\calP$ as input. Then, we report the
running time of {\ouralgo} using this partition. 
The local computations on the supernodes $G_i$
(lines~\ref{ln:frontiers} and~\ref{ln:dijkstra} of Alg.~\ref{algo:buildskeleton}) can be done in parallel across the $G_i$'s. 
Hence, the running time we report is the sum of: 
(i) the running time of \buildskeleton\ on the largest supernode $G_i$
$(ii)$ running {\brandessk} on the \skeleton\ and $(iii)$ computing the centrality of 
\emph{all} remaining nodes in $G$.

\spara{Implementation:} 
In all our experiments we compare the running times
of \ouralgo\ to \brandes~\cite{Brandes01} on weighted undirected graphs.  
While there are several high-quality implementations of \brandes\ available, we use here 
our own implementation of {\brandes} and, of course, {\ouralgo}. 
All the results reported here correspond to our Python implementations of both algorithms.
The reason for this is, that we want to ensure a fair comparison between the algorithms, where the algorithmic aspects of the running times are compared 
as opposed to differences due to more efficient memory handling, properties of the used language, etc. As the \brandessk\ algorithm run on the \skeleton\ is almost identical to \brandes\ (see Section~\ref{sec:algorithms}),
in our implementation we use the exact same codes for \brandes\ as {\brandessk}, except for appropriate changes that take into account edge multiplicities. We also make our code available\footnote{available at: \url{http://cs-people.bu.edu/edori/code.html}}.

\spara{Hardware:} All experiments were conducted on a
machine with Intel X5650 2.67GHz CPU  and 12GB of memory.

\spara{Datasets:} 
We use the following datasets:

\emph{\wikivote\ dataset~\cite{Leskovec2010}:} The nodes in this  graph correspond to users
and the edges to users' votes in the election to being promoted to certain
levels of Wikipedia adminship. We use the graph as undirected, assuming
that edges simply refer to the user's knowing each other.  The resulting
graph has $7066$ nodes and $103$K edges.

\emph{\asdata\ dataset:~\cite{Gill11}} The \asdata\ graph corresponds to a communication network of
who-talks-to whom from BGP logs. We used the directed Cyclops AS graph
from Dec. 2010 \cite{Gill11}.  The nodes represent
Autonomous Systems (AS), while the edges represent the existence of
communication relationship between two ASes and, as before, we assume the
connections being undirected.  The graph contains $37K$
nodes and $132$K edges, and has a power law degree distribution. 

\emph{\eumail\ dataset~\cite{Leskovec07}:} 
This graph represents email data from a European research
 institute. Nodes of the graph correspond to the senders and recipients of
 emails, and the edges to the emails themselves.  Two nodes in the graph
 are connected with an undirected edge if they have ever exchanged an
 email. 
 The 
graph has $265$K nodes and $365$K edges.

\emph{\dblp\ dataset~\cite{Yang12}:} The \dblp\ graph contains the
co- authorship network in the computer science community. Nodes correspond  to
authors and edges capture co-authorships. There are 
$317$K nodes and $1$M edges.

For all our real datasets, we 
pick $200$ nodes (uniformly at random) to form the target set $S$.

\spara{Graph-partitioning:} The speedup ratio of \ouralgo\  over \brandes\ is determined 
by the structure of the \skeleton ($\Gsk$) that is induced by the input graph partition.  

In practice, graphs that benefit most of \ouralgo\
are those that have small $k$-cuts, such as those that have distinct community structure.
On the other hand, graphs with large cuts, such as  power-law graphs do not benefit that much from applying the partitioning of \ouralgo.

For our experiments we partition the input graph into subgraphs using
well-established graph-partitioning algorithms, which aim to either
find densely-connected subgraphs or sparse cuts.  Algorithms with the former objective fall under \emph{modularity clustering}~\cite{Clauset04,Newman04,Radicchi04,Rotta11} while the latter 
are  \emph{normalized cut} algorithms~\cite{andersen10local,Dhillon07,Hendrickson95,kannan04clusterings,karypis98multilevel,ng01spectral,Schaeffer07}.
We choose the following three popular algorithms from these groups: \modularity, \graclus\ and {\metis}.

\emph{\modularity:} {\modularity} is a hierarchical agglomerative algorithm that uses the modularity
optimization function as a criterion for forming clusters. Due to the nature of the  
objective function, the algorithm decides the number of output clusters automatically and the number
of clusters need not be provided as part of the input.  {\modularity} is described in  Clauset {\it et al} \cite{Clauset04} and its 
implementation is available at: \url{http://cs.unm.edu/~aaron/research/fastmodularity.htm}

\emph{{\graclus}:} 
{\graclus} (\texttt{graclus}) is a normalized-cut partitioning algorithm that was first introduced by 
 Dhillon {\it et al.}~\cite{Dhillon07}. An implementation of {\graclus}, that uses a kernel $k$-means heuristic for producing a partition, is available at: \url{cs.utexas.edu/users/dml/Software/graclus.html}.
  {\graclus} takes as input $k$, an upper bound on the number of clusters
 of the output partition. For the rest of the discussion we will use
 {\graclus}-$k$ to denote the {\graclus} clustering into at most $k$ clusters.
 
 \emph{{\metis}:} This algorithm~\cite{karypis98multilevel} is perhaps the most widely used  normalized-cut partitioning algorithm. It does hierarchical graph bi-section with the objective  to find a balanced partition that minimizes the total edge cut between the different parts of the partition. An implementation of the algorithm is available at: \url{glaros.dtc.umn.edu/gkhome/views/metis}.
Similar to {\graclus}, {\metis} takes an upper bound on the number of clusters $k$ as part of the input. Again,
we use the notation {\metis}-$k$ to denote the {\metis} clustering into at most $k$ clusters.

We report
the running times of the three clustering algorithms in 
Table~\ref{tab:clustering-time}.  
Note that for both {\graclus}
and {\metis} their running times depend on the input number of clusters $k$ -- the larger the value
of $k$ the larger the running time. The table summarizes
the largest running time for each dataset (see Table~\ref{tab:thetable} for the value of $k$ for each dataset).
 Note that the running times of the clustering algorithms cannot be compared to the running time of \ouralgo\ for two reasons; these algorithms are implemented using a different (and more efficient) programming language than Python and are highly optimized for speed, while our implementation of \ouralgo\ is not. We report Table~\ref{tab:clustering-time} to compare the various clustering heuristics against each other.

\begin{table}
  \centering
  \topcaption{Running time (in seconds) of the clustering algorithms (reported for the largest number of clusters per algorithm) and of {\brandes} (last column).}
  \begin{tabular}{lccc|cc}
    \toprule
     & \modularity\ & \graclus\ & \metis\ & & {\brandes}\\
\midrule
\wikivote\ & $13$ & $1.29$&$1.9$ & & $5647$\\
\asdata\  &$6780$ & $256.58$ &$9.5$ & & $606$\\
\eumail\ &$1740$ &  2088&$12.2$ & & $14325$ \\
\dblp\ &$3600$ & $109.22$ &$5.5$ & & $28057$\\
\bottomrule
     \end{tabular}
    \label{tab:clustering-time}
\end{table}

\spara{Results:}
The properties of the partitions produced for our  datasets by the different
clustering algorithms, as well as the corresponding running times of {\ouralgo} for each
partition are shown in 
Table~\ref{tab:thetable}. In case of \graclus\ and \metis\ we experimented with several (about $10$) 
values of $k$. We report  for three different values of $k$ (one small, one medium and one large) for each dataset. The values were chosen in such a way, that the $k$-clustering with the best results in \ouralgo\ is among those reported. As a reference point, we report in Table~\ref{tab:clustering-time} the running times of the original \brandes\ algorithm on our datasets.

\begin{table*}
  \centering
  \topcaption{Properties of the partitions ($N$: number of frontier nodes in {\skeleton}, $M$: number of edges in {\skeleton}, $k$: number of supernodes, LCS: number of original nodes in the largest cluster) produced by different clustering algorithms
  and running time of {\ouralgo} on the different datasets.}
  \begin{tabular}{lccccccc}
    \toprule
\multicolumn{8}{c}{{\wikivote} dataset}\\
    \toprule    
    & \modularity\ & {\graclus}-$100$ & {\graclus}-$1K$ &
  {\graclus}-$2K$ & {\metis}-$100$ & {\metis}-$1K$& {\metis}-$2K$ \\
 \midrule
$N$ &$3833$ & $5860$ & $6365$& $6432$ & $4920$& $5854$ & $6181$\\
$M$  & $26147$& $89318$ & $96111$ & $96743$ & $91545$& $92091$&
$97270$ \\
$k$  & $29$ & $100$&$1000$  &$2000$& $98$& $989$& $1927$\\
LCS  &$3059$ & $172$ &$12$ & $10$ & $258$ & $496$& $50$  \\
\toprule
\multicolumn{8}{c}{{\ouralgo} running time in seconds}\\
\midrule
  & $209.43$ & $75.27$  & $90.23$  & $96.65$ & $71.59$ & $73$ & $77.71$\\
\bottomrule
\multicolumn{8}{c}{\asdata\ dataset} \\
      \toprule
     & \modularity\ & {\graclus}-$1K$ & {\graclus}-$10K$ &
  {\graclus}-$15K$ & {\metis}-$1K$ & {\metis}-$10K$& {\metis}-$15K$ \\
\midrule
$N$ &$14104$ & $31139$ &$34008$ & $34418$ & $25022$& $32572$ & $35244$\\
$M$  & $28815$& $111584$ & $120626$ & $121833$ & $93989$& $117808$& $125556$\\
$k$  & $156$& $1000$&$10000$ &$15000$ &$991$ &$9966$ & $14484$\\
LCS  &$8910$ &$732$ & $10$ & $10$  &$1304$ &$433$ & $18$ \\
\toprule
\multicolumn{8}{c}{{\ouralgo} running time in seconds}\\
\midrule
 &$1666$ & $430.97$  & $447.97$ & $486.91$ & $417$ & $420.93$ & $458.71$ \\
\bottomrule
\multicolumn{8}{c}{\eumail\ dataset}\\
    \toprule    
     & \modularity\ & {\graclus}-$1K$ & {\graclus}-$3K$ &
  {\graclus}-$5K$ & {\metis}-$1K$ & {\metis}-$3K$& {\metis}-$5K$ \\
\midrule
$N$ &$19332$ &$143636$ & $208416$&$215397$ & $42333$& $54249$ & $50171$\\
$M$  & $45296$& $231089$ &  $208416$&  $215397$& $117147$& $132573$&
$129071$ \\
$k$  & $45296$& $231089$ & $319006$& $327263$& $995$ &$2996$ & $4996$\\
LCS  & $53224$& $7634$& $7633$  & $7633$ & $8917$& $7407$ & $7271$\\
\toprule
\multicolumn {8}{c}{{\ouralgo} running time in seconds}\\
\midrule  
& $188.79$ & $5670$ & $7816.7$& 8291.3& $3601$ & $2101$ & $1872$ \\
\bottomrule
  \multicolumn{8}{c}{{\dblp} dataset}\\
    \toprule
     & \modularity\ & {\graclus}-$100$ & {\graclus}-$1K$ &
 {\graclus}-$5K$ & {\metis}-$100$ & {\metis}-$1K$& {\metis}-$5K$ \\
\midrule
$N$ &$102349$ & $98281$ & $130643$ & $141955$ & $104809$& $119417$ & $132661$\\
$M$  & $146584$& $164989$ & $267350$ & $310472$ & $203834$& $257383$&
$318969$ \\
$k$  & $3203$& $100$ & $1000$&$5000$& $100$ & $1000$& $4999$\\
LCS  & $55897$&$116252$  &$26666$ &$21368$ & $3270$ & $344$ & $93$ \\
\toprule
\multicolumn{8}{c}{{\ouralgo} running time in seconds}\\
\midrule
 & $3600$   & $95982$  & $16405$ & $10335$ & $13574$ & $5805$ &  $5709$ \\
\bottomrule
     \end{tabular}
    \label{tab:thetable}
\end{table*}

In Table~\ref{tab:thetable} $N$ and $M$ refer to the number of nodes and edges in the \skeleton. Remember, that the set of nodes in the  skeleton is the union of the frontier nodes in each supernode.  Hence, $N$ is equal to the total number of frontier nodes induced by $\calP$.   Across datasets we can see  quite similar values, depending on the number of clusters used. \modularity\ seems to yield the lowest values of $N$ and $M$. 
 The third and fourth rows in the table contain the number of clusters $k$
 (the size of the partition) for each algorithm and the total number of nodes  (from the input graph) 
 in  the largest
 cluster of each partition.

 The ultimate measure of performance is the running time of \ouralgo\ in
 the last row of the table. We compare the running times of \ouralgo\ to the running time
 of \brandes\ in Table~\ref{tab:clustering-time} -- last column.
On the \wikivote\ data {\brandes} needs 5647 seconds  while the corresponding time
for {\ouralgo} can be as small as 72 seconds! Note that the best running time for this dataset is achieved using the {\metis}-$100$ partition. Suggesting that the underlying "true" structure of the dataset consists of approximately $100$  communities of Wikipedia users. High speedup ratios are also achieved on {\eumail} and {\dblp}. For those {\brandes} takes 14325 and 28057 seconds respectively,
while the running time of {\ouralgo} can be 189 and 3600 seconds respectively. This is
an 8-fold speedup on \dblp\ and 75-fold on \eumail. 

If we compare the running time of \ouralgo\ applied to the different
 partitions, we see that  the algorithm with input  by \metis\ is
 consistently faster than the same-sized partitions of \graclus.  Further, 
  on \eumail\ and \dblp\  \ouralgo\ is the fastest with the \modularity\
 partition. Note the size of $\Gsk$ for each of these datasets. In case
 of \eumail\ \modularity\ yields a skeleton where $N$ is only $7\%$ of the
 original number of nodes and $M$ is $12\%$  of the edges. The
 corresponding rations on \dblp\ are $32\%$ and $14\%$. This is not surprising, as both datasets are known for their
 distinctive community structure, which is what \modularity\ optimizes
 for. 
For {\asdata}, {\ouralgo} exhibits again smaller running time than {\brandes}, yet the improvement
is not as impressive. Our conjecture is that this dataset does not have an inherent clustering structure
and therefore {\ouralgo} cannot benefit from the partitioning of the data.

 Note that the running times we report here refer only to the execution time of
 {\ouralgo} and do not include the actual time required for doing the clustering --
 the running times for clustering are reported in 
 in Table~\ref{tab:clustering-time}. However, since the preprocessing has to be done only once and the space increase is only a constant  factor, \ouralgo\ is clearly of huge benefit.

{\small
\bibliographystyle{abbrv}
\bibliography{betweenness}  
}

\end{document}